\documentclass[aps,prb,showpacs]{revtex4}

\usepackage{graphicx}
\usepackage{dcolumn}
\usepackage{bm}
\usepackage{amsmath}
\usepackage{color}

\begin{document}

\title{Magnetic phase stability of monolayers: Fe on
Ta$_{x}$W$_{1-x}$(001) random alloy as a case study}



\author{M. Ondr\'a\v{c}ek}\email{ondracek@fzu.cz}
\affiliation{Institute of Physics ASCR, Cukrovarnick\'a 10, CZ-162 00
Praha 6, Czech Republic}

\author{O. Bengone}
\affiliation{IPCMS, BP43, 23, rue du Loess, F-67034 Strasbourg Cedex 2, 
             France}

\author{J. Kudrnovsk\'y}
\affiliation{Institute of Physics ASCR, Na Slovance 2, CZ-182 21
Praha 8, Czech Republic}

\author{V. Drchal}
\affiliation{Institute of Physics ASCR, Na Slovance 2, CZ-182 21
Praha 8, Czech Republic}

\author{F. M\'aca}
\affiliation{Institute of Physics ASCR, Na Slovance 2, CZ-182 21
Praha 8, Czech Republic}

\author{I. Turek}
\affiliation{Institute of Physics of Materials ASCR, 
\v{Z}i\v{z}kova 22, CZ-616 62 Brno, Czech Republic}

\date{\today}

\begin{abstract}

We present a new approach to study the magnetic phase stability
of magnetic overlayers on nonmagnetic substrates.
The exchange integrals among magnetic atoms in the overlayer
are estimated in the framework of the adiabatic approximation
and used to construct the effective classical two-dimensional Heisenberg
Hamiltonian.
Its stability is then studied with respect to a large number of
collinear and non-collinear magnetic arrangements 
which include, as special cases, not only
ferromagnetic and various antiferromagnetic configurations, but
also possible incommensurate spin-spiral structures.
This allows us to investigate a broader class of systems than a
conventional total energy search based on few, subjectively 
chosen configurations.
As a case study we consider the Fe-monolayer on the random
nonmagnetic bcc-Ta$_{x}$W$_{1-x}$(001) surface which was
studied recently by a conventional approach.
We have found a crossover of the ground state of the Fe 
monolayer from the ferromagnet on the Ta surface to the 
$c(2\times2)$ antiferromagnet on the W surface and that 
at the composition with about 20 \% of Ta an incommensurate 
magnetic configuration might exist.

\end{abstract}



\maketitle

\section{Introduction}

A deeper understanding of the magnetic ground state 
as well as of finite-temperature properties of a system 
with local magnetic moments can be obtained in terms of 
corresponding exchange interactions.
This is particularly true for new, artificially prepared systems,
like, e.g., monolayers of magnetic atoms on nonmagnetic substrates.
A strong hybridization of the electronic states of the magnetic 
atoms and of the substrate may significantly influence the magnetic 
ground state of these systems.
We mention as a typical example the Fe-monolayer on bcc-W(001)
substrate.
While bcc-Fe in bulk state is a ferromagnet (FM),
the Fe-monolayer on the bcc-W(001) surprisingly has a $c(2\times 2)$
antiferromagnetic (AFM) ground state, as was recently found by 
spin-polarized STM measurements \cite{afm-few}.

In a subsequent theoretical study, 
Ferriani {\it et al.}\cite{prl-stefan} demonstrated a strong dependence of
the magnetic state of an Fe-monolayer on the substrate by studying its stability 
on the (001)-face of bcc-Ta$_{x}$W$_{1-x}$ random substrates.
It was found that the ground state of the Fe-monolayer on 
Ta(001) is the FM rather than the AFM one and there thus exists 
an interesting crossover between these two magnetic configurations 
at an intermediate substrate composition.
Results of this study also confirm the robustness of such
behavior - even the unrelaxed system exhibits the same behavior
although large inward relaxation of about 18~\% exists between
the Fe monolayer and the first substrate W-layer,
which can modify the result quantitatively. 
The authors of Ref.~\onlinecite{prl-stefan} also clearly
demonstrated the usefulness of exchange interactions for
a deeper understanding of the phenomenon.

Based on these facts we wish to present here a more general 
approach to the study of the magnetic stability of such systems.
Let us first remind the reader of the approach adopted in paper \cite{prl-stefan}
as well as in a number of other studies.\cite{r_2002_kbb, r_2002_sh, 
r_2002_kht, r_2005_nr, r_2008_smo}
Exchange interactions are estimated by evaluating the total 
energies of the FM and a few AFM configurations by conventional 
bandstructure methods.
In particular, in the case of Fe overlayer on the W(001) or
Ta(001), the authors of Ref.~\onlinecite{prl-stefan} 
chose $c(2\times 2)$- and $p(2\times 1)$-AFM 
configurations and mapped the total energy differences of 
these configurations and of the reference FM state onto the classical 
two-dimensional Heisenberg Hamiltonian (HH).
In this way they obtained two nearest-neighbor exchange couplings.
This approach has certain limitations: 
(i) the choice of AFM configurations is dictated by requirements of simplicity,
although any other configuration can be used as well.
One has to be aware that if some atypical configuration is chosen,
the results can be biased;
(ii) the number of determined exchange interactions depends
on the number of chosen configurations, which is limited.
This is in a striking contrast with the existence of long-range
interactions in such systems.
For example, the presence of the surface state, e.g., on the
fcc(111) surfaces of noble metals leads to the decay of
corresponding exchange interactions with the distance $d$ as
$d^{-2}$ (see Ref.~\onlinecite{lauko,hyldg}) as compared to
the $d^{-3}$-decay known from bulk systems (the RKKY-interactions).
In addition, the bandstructure methods do not yield directly individual
exchange interactions, but rather their certain infinite sums;
(iii) the use of conventional bandstructure methods is strongly
complicated in the presence of disorder, either in the magnetic
overlayer or in the substrate.
On the other hand, if the substrate disorder is weak, a simple approach
(the virtual-crystal approximation) can be successfully used as it was
done in Ref.~\onlinecite{prl-stefan} for the case of Fe-overlayer on
the TaW-alloy substrate.

An alternative approach is to estimate magnetic interactions 
among magnetic atoms directly by evaluating the energy 
cost related to small rotations of spins at 
arbitarily chosen two sites with the help of the adiabatic 
theorem and the Green function method.
This approach, first introduced by Lichtenstein \emph{et al.} \cite{lie}
for bulk ferromagnets and by Oguchi \emph{et al.} \cite{oguchi} 
for bulk paramagnets within the disordered local moment (DLM) state, 
was then successfully generalized to 
a number of magnetic systems including magnetic overlayers 
on crystal substrates \cite{pajda} (for further magnetic 
systems see the review paper \cite{eirev}).
In this method the limitations mentioned above are removed.
As the method is formulated in the real space, the systems without 
perfect translational symmetry (e.g. random alloys, overlayers, and
surfaces) can be treated on an equal footing with the crystals.

One remark concerns the choice of the reference state
from which the exchange interactions are extracted.
Conventionally, this is the FM state \cite{lie,pajda,pajda2}, but it is
possible to employ also the AFM reference state or any other 
reference state such as the DLM state.
In the DLM state the spin-spin correlations are missing while 
in the FM- or AFM-states the spins are strongly correlated.
The DLM state is thus particularly suitable for discussion of  
the electronic origin of the magnetic phase stability as no 
specific magnetic order is assumed in this reference state.
A traditional approach to the exchange interactions in
the DLM-reference state \cite{oguchi, r_1998_su, r_2004_rss}
rests on the generalized perturbation method (GPM); \cite{r_1991_bg}
however, the final formula can also be obtained from the method
of infinitesimal rotations of local moments applied to a random
binary alloy.
This equivalence is briefly sketched in Section II and in the
Appendix.

Once the exchange interactions are found, and it should be
noted that they can be found even for very distant pairs
of spins, one can study the ground state of the classical HH.
In this way one can search over much broader class of
possible magnetic systems, including incommensurate
configurations, than it is possible by using a conventional
supercell approach.
It should be noted, however, that the supercell approach 
can be highly accurate and useful and we wish to point out
the usefulness and complementarity of both above mentioned
approaches for various aspects of the magnetic stability 
rather than to argue that one approach is better as compared 
to the other.

\section{Formalism}

The electronic properties of the Fe-overlayer on random 
Ta$_{x}$W$_{1-x}$(001) substrate will be studied in the
framework of the tight-binding linear muffin-tin orbital
(TB-LMTO) method combined with the coherent potential
approximation (CPA) to treat alloy disorder in the substrate
as well as the magnetic disorder in the DLM state.
The effect of the surface is included in the framework 
of the surface Green function (SGF) approach \cite{book} which employs a
realistic semi-infinite sample geometry (no slabs or periodic
supercells).
The one-electron potential is treated within the atomic
sphere approximation, but the dipole barrier due to the
redistribution of electrons in the vicinity of the surface 
is included in the formalism.
The TB-LMTO-SGF method can include the effect of layer
relaxations approximately provided that they are known either 
from high-accuracy ab-initio calculations or from experiment.

An important advantage of the TB-LMTO-SGF approach is the
possibility to estimate exchange interactions between
magnetic atoms in the overlayer.
We will describe such magnetic interactions in terms of 
the two-dimensional classical Heisenberg Hamiltonian,
\begin{equation}
\label{HH-def}
H = -{\sum_{i \neq j}} J_{ij} \, {\bf e}_i \cdot {\bf e}_j \, ,
\end{equation}
in which $J_{ij}$ denotes the exchange integral between 
Fe-spins at sites $i$ and $j$ on the surface, and $ {\bf e}_i$ and 
${\bf e}_j$ are unit vectors in the directions of the local 
magnetization on sites $i$ and $j$, respectively.

Exchange interactions $J_{ij}$ can be evaluated in the 
the FM reference state and in the framework of the TB-LMTO-SGF method as
\begin{equation}\label{Jij-fm}
J_{ij} = \frac{1}{4\pi} \, \mathrm{Im} \, \int_{C}  \mathrm {tr}_{L} 
\{ \Delta_{i}(z) \, \bar{g}^{\uparrow}_{ij}(z) \, 
\Delta_{j}(z) \, \bar{g}^{\downarrow}_{ji}(z) \}
\, \mathrm{d}z \, ,
\end{equation}
where $\Delta_{i}(z)$ characterizes the exchange splitting of
the Fe-atom at the site $i$, $\bar{g}^{\sigma}_{ij}(z)$  is 
the configurationally averaged Green function describing the motion of 
an electron between Fe-sites $i$ and $j$ in the magnetic overlayer on a 
random nonmagnetic substrate which corresponds to the spin $\sigma$,
$\sigma=(\uparrow, \downarrow)$, and the integration is done over 
the contour $C$ in the complex energy plane $z$ which starts below the 
valence band and ends at the Fermi energy.
Symbol $\mathrm{tr}_L$ denotes the trace over the atomic orbitals 
($L\equiv (l,m)$).
The quantity $\Delta_{i}(z)$ is defined in the 
TB-LMTO method in terms of the potential functions
$P_{i}^{\,\sigma}(z)$ of Fe atoms as 
$\Delta_{i}(z)=P^{ \uparrow}_{i}(z)-
P^{ \downarrow}_{i}(z)$.
The above expression represents a straightforward generalization of
the corresponding bulk expression \cite{lie,eirev} since it is 
formulated in the real space.
We refer the reader to Refs.~\onlinecite{surf1,surf2, book} for details 
of evaluation of the real-space configurationally averaged Green 
function $\bar{g}^{\sigma}_{ij}(z)$ for the magnetic 
overlayer on a random substrate. 

In the DLM-reference state, the exchange interactions
$J_{ij}$ bear a standard form of effective pair
interactions of the GPM. They are given explicitly by
\begin{equation}\label{Jij-dlm}
J_{ij} = \frac{1}{4\pi} \,
\mathrm{Im} \int_{C} \mathrm {tr}_{L}
\left\{ \tau_{i}(z) \, \bar{g}_{ij}(z) \,
\tau_{j}(z) \, \bar{g}_{ji}(z)
\right\} \mathrm{d}z \, ,
\end{equation}
where the $\bar{g}_{ij}(z)$ is
the spin-independent configurationally averaged Green function
in the DLM state between Fe-sites $i$ and $j$.
The electronic structure of the DLM state is treated in an alloy
analogy, i.e., the Fe-sites are randomly occupied by two Fe-species
with equal probability, representing local Fe-moments pointing in
two opposite directions.
The quantity $\tau_{i}(z)$ is defined as $\tau_{i}(z)
 = t^{\uparrow}_{i}(z) - t^{\downarrow}_{i}(z)$,
where the symbols $t^{ \sigma}_{i}(z)$
($\sigma = \uparrow, \downarrow$) refer to single-site t-matrices
of the TB-LMTO method describing scattering due to both magnetic
species on the $i$-th site with respect to a spin-independent
effective DLM medium, see Eq.~(\ref{eqa_tmat}).
In the Appendix, the form of Eq.~(\ref{Jij-dlm}) is
derived from the previous formula, Eq.~(\ref{Jij-fm}),
applied to the DLM state within the alloy analogy.

The exchange integrals $J_{ij}$ characterize magnetic 
interactions between two particular atomic sites.
It is also convenient to define the exchange parameter $J_{0}$ 
that reflects the molecular field experienced by a single moment 
in the ferromagnetic layer
\begin{equation}\label{J0-sum}
J_{0} = \sum_{i \neq 0} J_{0i} \, .
\end{equation}

A negative value of this parameter indicates an instability
of the FM state.
A more detailed analysis of the magnetic stability employs the
lattice Fourier transform of the pair exchange interactions
$J_{ij}$, see Section III C; the parameter
$J_{0}$ represents a special case of the Fourier transformed
interaction, namely, its value for zero reciprocal vector, see
Eq.~(\ref{Jq-def}).

\section{Results and discussion}

The calculations were done assuming the Vegard's law for the bulk
lattice constant $a$ of bcc-Ta$_{x}$W$_{1-x}$ random alloy
($a_{\rm Ta}=3.300~{\rm \AA}$ and $a_{\rm W}=3.165~{\rm \AA}$).
The electronic structure is calculated selfconsistently for a
model of the system which consists of 7 atomic layers representing
the alloy substrate, one Fe layer, and 4 layers of empty spheres
representing vacuum that are embedded between the semi-infinite
substrate and the semi-infinite vacuum.
Similarly as in Ref.~\onlinecite{prl-stefan}, we assume a constant
inward layer relaxation of 18~\% between the Fe-overlayer and the 
bcc(001)-Ta$_{x}$W$_{1-x}$ substrate.
We have employed the ${\it spdf}$-basis and the Vosko-Wilk-Nusair 
exchange-correlation LSDA potential in all calculations.
We used 20 points on the contour in the complex energy plane and 
210 k-points in the irreducible wedge of the surface Brillouin zone 
in selfconsistent calculations and up to 19600 k-points in the 
surface Brillouin zone to calculate exchange integrals for 172 
inequivalent neighbors.

\subsection{Density of states}

\begin{figure}
\begin{center}
\includegraphics[width=9cm]{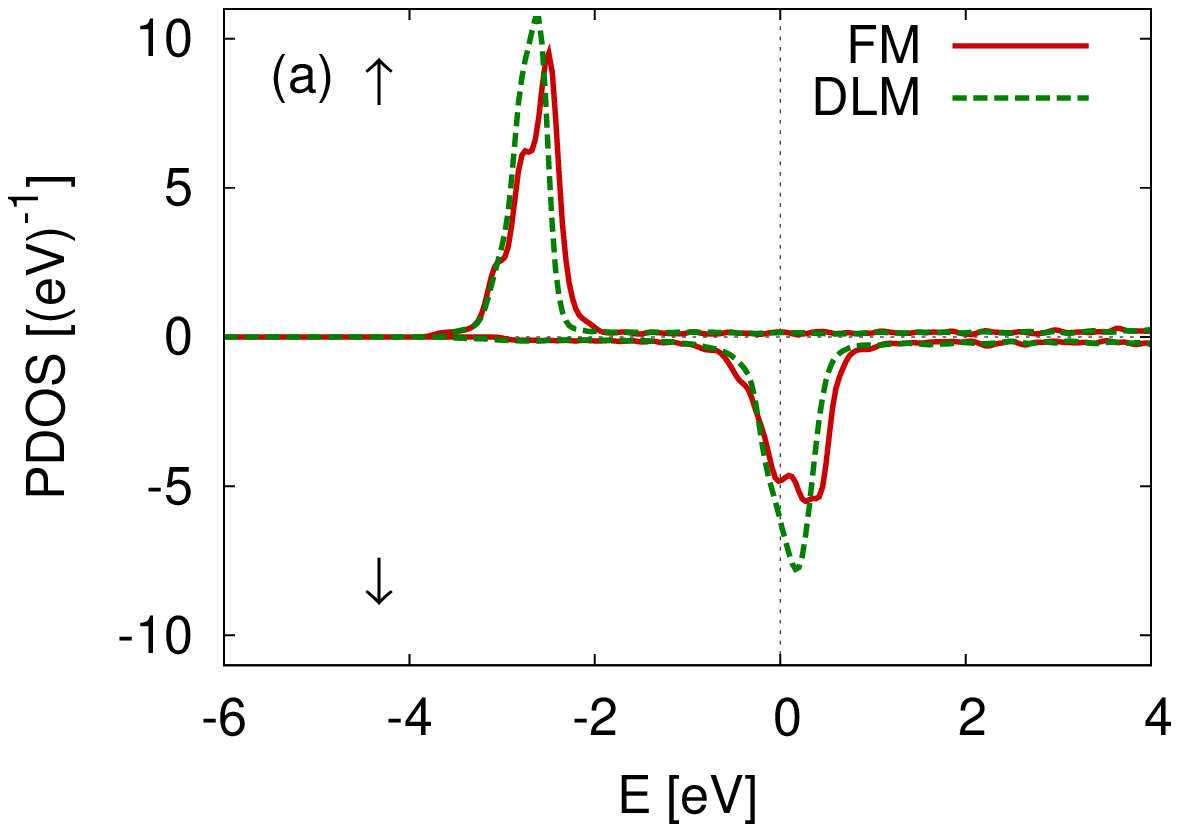}\\[4mm]
\includegraphics[width=9cm]{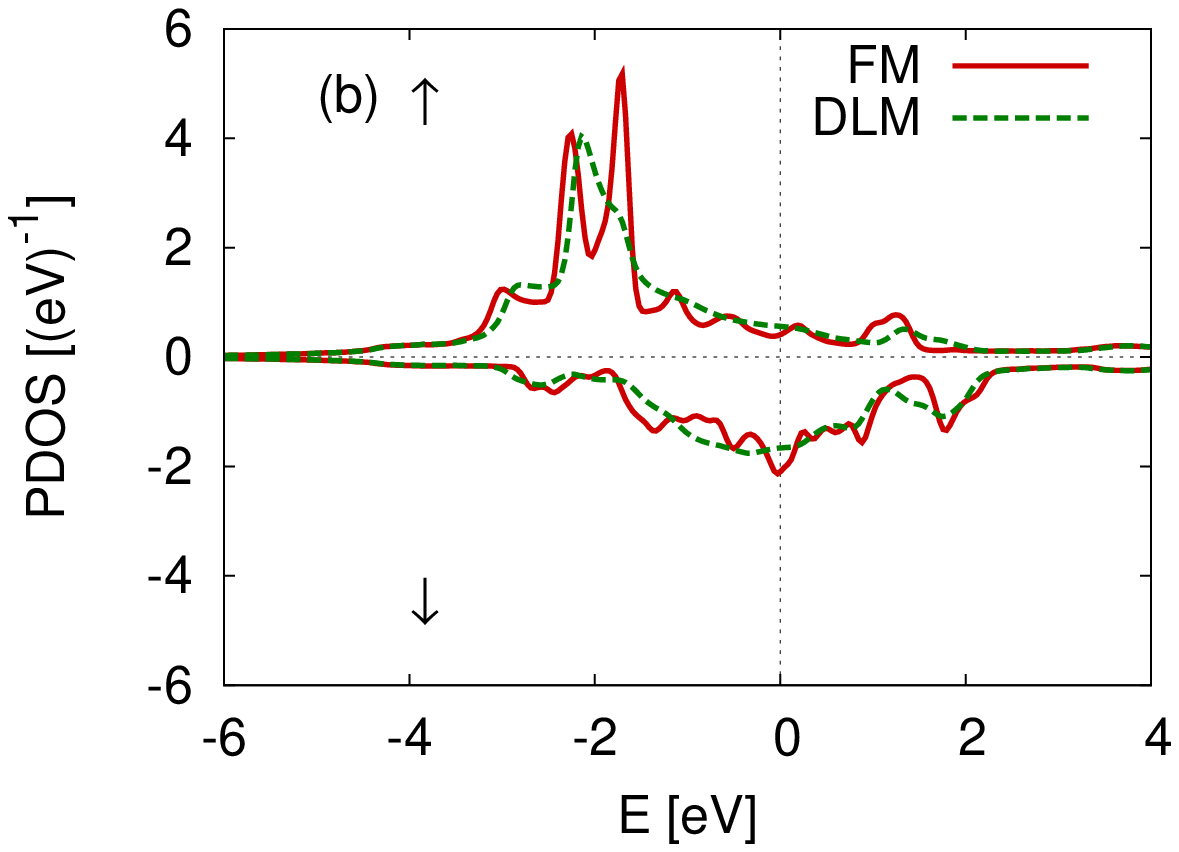}\\[4mm]
\includegraphics[width=9cm]{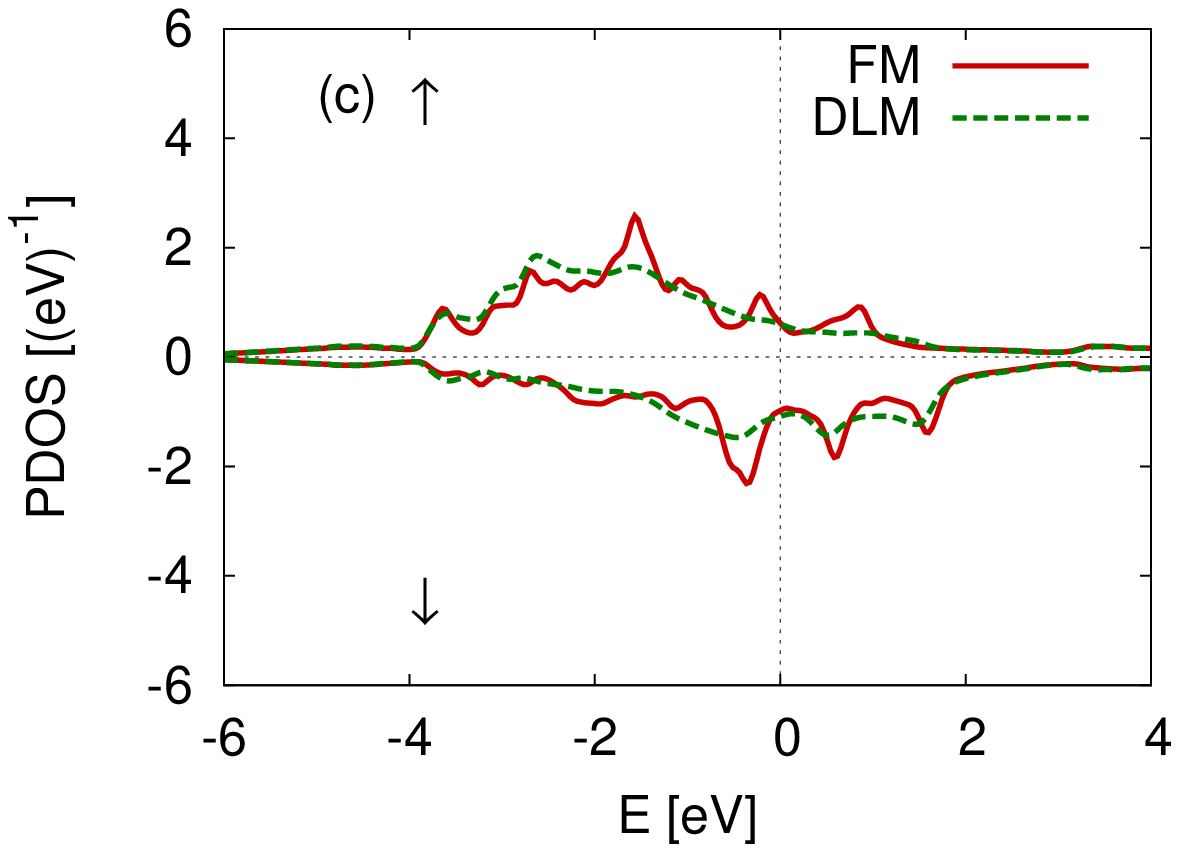}
\end{center}
\caption{\label{dos} (color online).
Spin-resolved densities of states projected on Fe-atoms for various
systems in the ferromagnetic and DLM state:
(a) an unsupported Fe monolayer, (b) Fe monolayer on the top of 
Ta(001) surface, and (c) Fe monolayer on the top of the W(001) surface.
The majority (minority) spin states are shown in the upper
(lower) panels.
The vertical lines denote positions of the Fermi levels.}
\end{figure}

We compare in Fig.~\ref{dos} the Fe-projected densities of states
(PDOS) for an unsupported monoatomic Fe(001) layer (assuming the in-plane lattice constant
of bcc-W, but the results for that of bcc-Ta are very similar)
with the PDOS for Fe$/$W(001) and Fe$/$Ta(001).
Results for the FM and for the DLM states are shown.
We observe a much stronger effect of hybridization among the Fe-located 
and the substrate states in the case of the W substrate than for
the Ta substrate.
The PDOS for Fe on the Ta substrate, especially that for the majority spin, 
is only weakly broadened as compared to the unsupported Fe layer
in striking constrast to the case of the W substrate.
We thus argue that the strong Fe-W hybridization is the reason 
for the AFM ordering on the W-substrate: the ground state
of the weakly hybridized Fe$/$Ta(001) is similar to that of 
the unsupported Fe layer, namely the FM state.
Roughly speaking, the stronger Fe-substrate hybridization also
reduces the DOS at the Fermi energy in the non-magnetic state and thus reduces the 
tendency to ferromagnetism (the Stoner criterion).
It should be noted that the generally broader minority PDOS as
compared to the majority one is due to the large exchange
splitting of Fe electron levels: majority levels are closer to the
nucleus and thus are more strongly bound giving a narrower
band and just the opposite holds for the more loosely bound minority bands.

\begin{figure}
\begin{center}
\includegraphics[width=12cm]{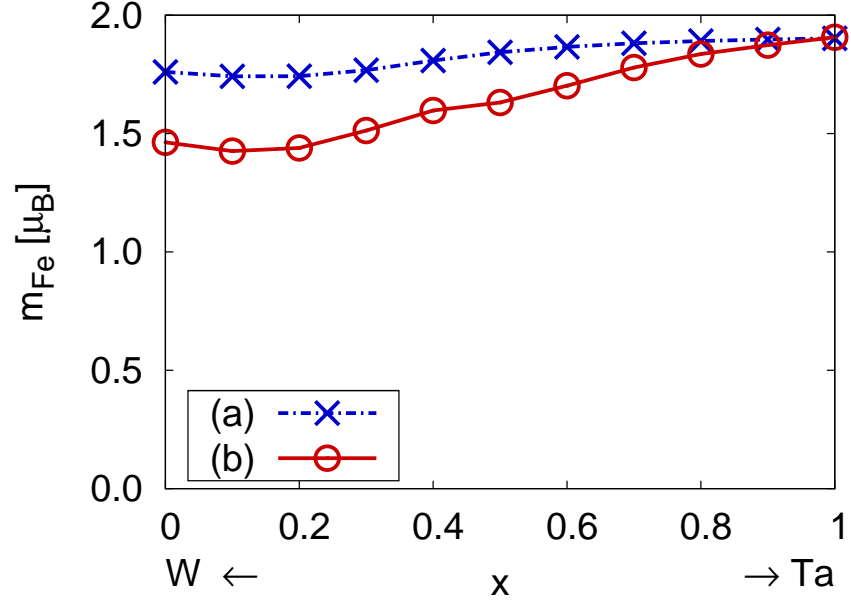}
\end{center}
\caption{\label{magmom} (color online).
The local magnetic moments $m_{\rm Fe}$ of Fe atoms in the
monolayer on Ta$_x$W$_{1-x}$(001) substrate as a function of the
alloy composition: (a) the DLM configuration (crosses,
$|m_{\rm Fe}|$ is shown) and (b) the FM configuration
(circles).
}
\end{figure}

Another indication of a stronger hybridization of the Fe-W states
as compared to the Fe-Ta states is documented by Fig.~\ref{magmom},
where the concentration dependence of local magnetic Fe-moments 
evaluated in the FM and DLM states is shown.
Generally, a stronger hybridization of Fe-W states results in an
increasing difference between FM and DLM magnetic moments.
The local magnetic Fe moments for the unsupported Fe layer are
large and almost the same (about 3.35~$\mu_{\rm B}$) in
both the FM and DLM configurations.
A stronger hybridization reduces more the local magnetic moments 
for the Fe$/$W(001) overlayer as compared to the Fe$/$Ta(001) one.

\begin{figure}
\begin{center}
\includegraphics[width=9cm]{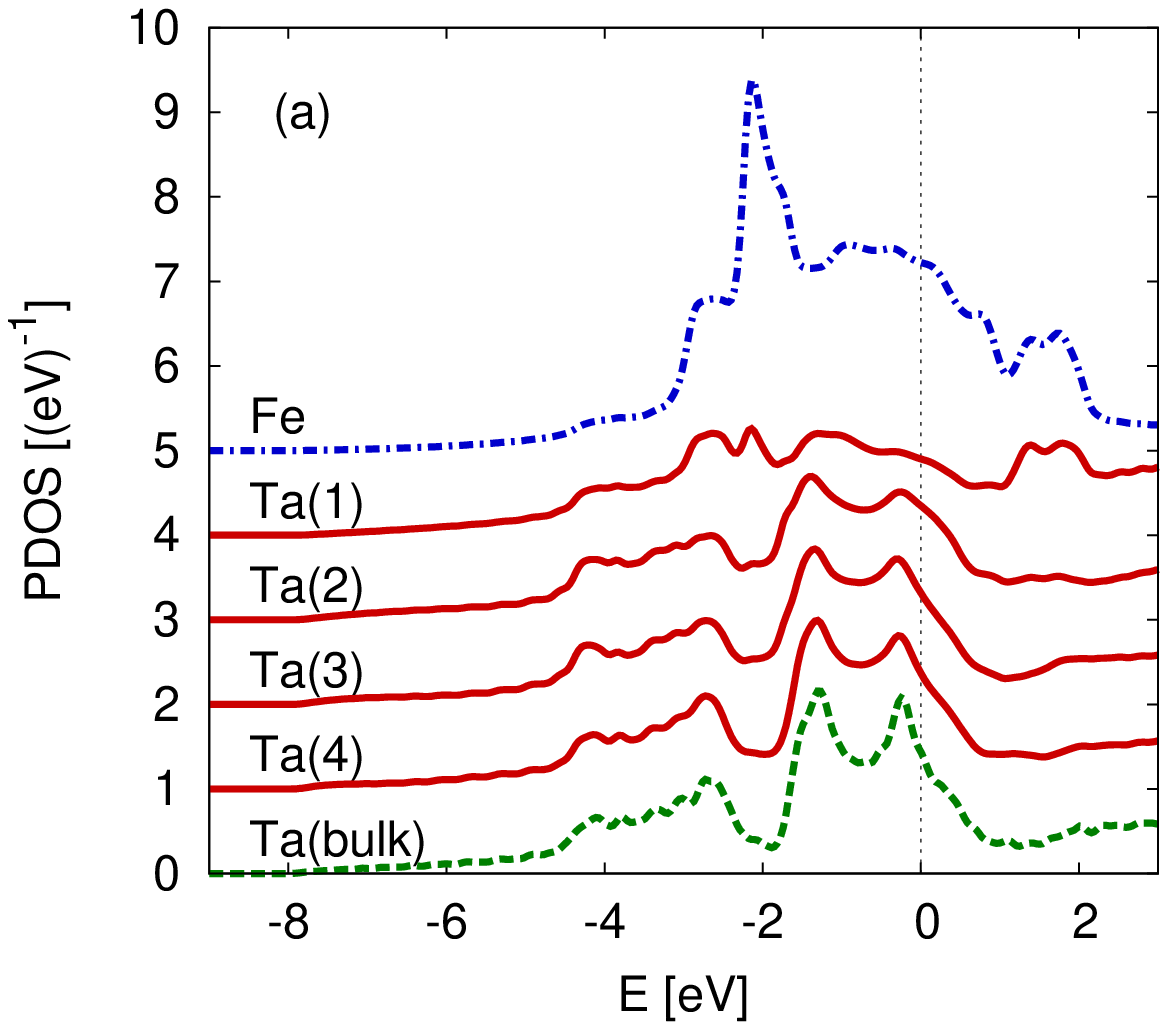}
\includegraphics[width=9cm]{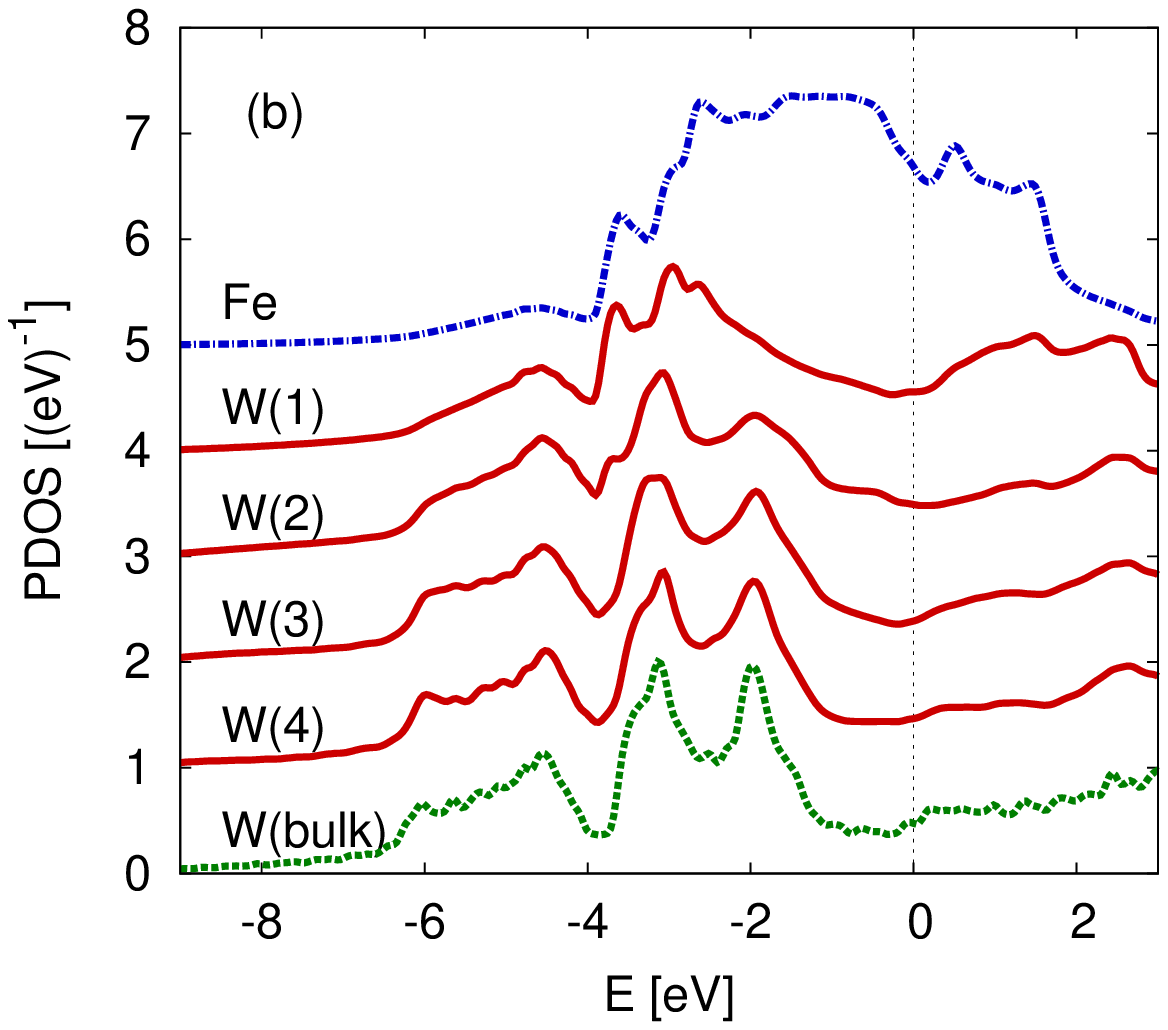}
\end{center}
\caption{\label{deepdos} (color online).
Layer-resolved densites of states averaged over spins for the 
Fe-overlayer in the disordered local moment state (dotted-dashed line), 
first four substrate layers (full line),
and for the bulk substrate (dashed line): (a) Fe$/$Ta(001) system, and 
(b) Fe$/$W(001) system. The curves are shifted vertically with respect 
to each other. Note the different scale for (a) and (b) plots.
The vertical lines denote positions of the bulk Fermi levels.
}
\end{figure}

In Fig.~\ref{deepdos} we present layer-resolved PDOS for
first few top layers of the Fe$/$W(001) and Fe$/$Ta(001) systems
(averaged over spins) and corresponding bulk substrates.
We first note a large similarity between the bulk Ta and W
total DOS which justifies the virtual crystal approximation
used in Ref.~\onlinecite{prl-stefan} to study the effect of
alloy disorder in the substrate.
The main difference is the position of the Fermi energy which
is shifted downwards in the bcc-Ta because of a smaller number
of valence electrons.
As a consequence, the Fermi energy lies inside the bonding
peak of the bcc-DOS while it is shifted to the energy region
between the bonding and antibonding states in the bcc-W.
We also mention that only the interface substrate layer and
to some extent also the second substrate layer differ noticeably
from the bulk substrate while the DOSs of other substrate layers 
are already quite similar to those of bulk.
This is, of course, a typical feature of metallic systems in
which the corresponding Friedel-like oscillations (which
originate from the presence of an abrupt change of the
charge density at the surface) are quickly damped inside
the substrate due to the high density of screening electrons.

\subsection{Exchange integrals}

\begin{figure}
\begin{center}
\includegraphics[width=12cm]{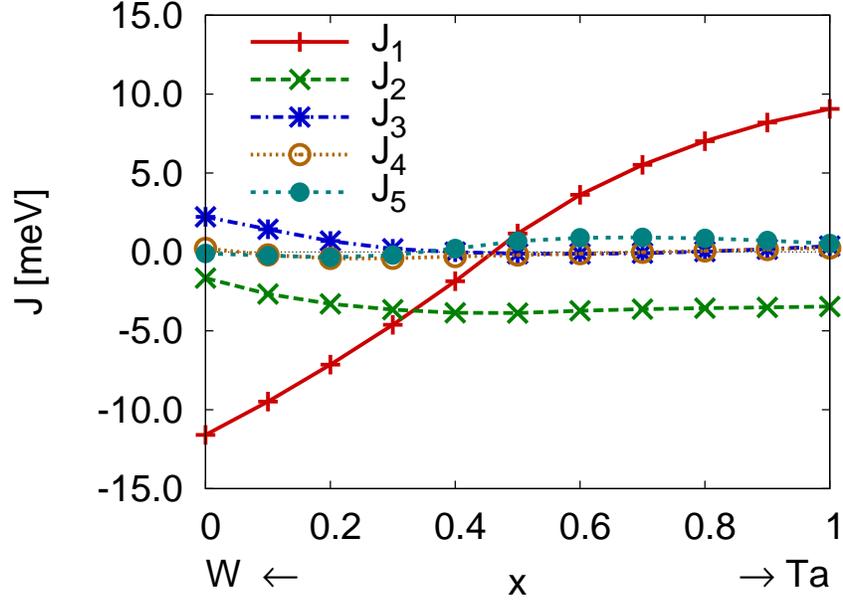}
\end{center}
\caption{\label{j5} (color online).
Exchange integrals between Fe-atoms (up to the fifth shell) in the 
Fe overlayer on the Ta$_{x}$W$_{1-x}$ (001) random alloy surface
as a function of the substrate alloy composition.}
\end{figure}

Concentration dependence of the exchange integrals $J_{s}$
for the first few shells of neighboring atoms $s=1-5$ is shown in Fig.~\ref{j5}.
First, this figure illustrates the principal difference between
the approach used in Ref.~\onlinecite{prl-stefan} and the present approach.
In the former method, the total enegy differences between three magnetic configurations,
namely the FM, $c(2\times 2)$- and $p(2\times 1)$-AFM configurations,
are used to estimate two exchange integrals.
One can call this approach the first-principle fitting as compared
to the direct first-principle calculation of exchange 
integrals in our present approach.
Clearly, realistic exchange integrals are not limited to the first
two shells and, in particular, not in the two-dimensional case where 
exchange integrals can decay in general more slowly than in the three 
dimensional case (see e.g. Ref.~\onlinecite{lauko,hyldg}).
It should be noted that we have plotted the exchange integrals without
corresponding degeneracies (multiplicities of equivalent surface atoms in the shells)
that are 4 or 8 in the present case.

The most remarkable feature observed in  Fig.~\ref{j5} is a dramatic 
change of the leading first nearest-neighbor (NN) interaction from 
the strongly AFM coupling in Fe$/$W(001) to the FM coupling in Fe$/$Ta(001).
The crossover between the AFM and FM couplings roughly coincides with 
the crossover between the FM and AFM ground state of the Fe overlayer.
The concentration  dependence of other interactions on the substrate 
alloy composition is generally weak.
The second NN interaction has the AFM character in the whole concentration 
range while the third NN interaction is FM-like for a W-rich substrate.
The first two NN interactions agree reasonably well with those
fitted from total energies for three magnetic configurations 
\cite{prl-stefan} both qualitatively and quantitatively (note
that the definitions of the Heisenberg Hamiltonian used in the present paper 
and in Ref.~\onlinecite{prl-stefan} differ by a factor of two).
The quantitative agreement of these NN interactions also proves 
that the chemical disorder in the TaW alloy substrate,
which Ref.~\onlinecite{prl-stefan} treated in a simple virtual crystal approximation
but which the present study treats in the CPA, is really weak.

\begin{figure}
\begin{center}
\includegraphics[width=12cm]{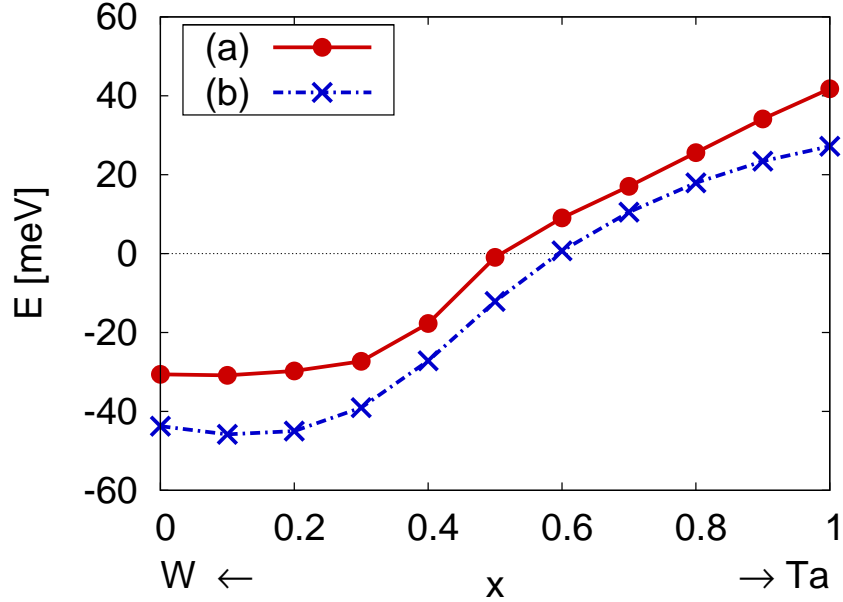}
\end{center}
\caption{\label{j0} (color online).
(a) The total energy difference of the disordered local moment (DLM) 
and ferromagnetic (FM) configurations per surface atom
($E_{\mathrm{tot}}^{\mathrm{DLM}}-E_{\mathrm{tot}}^{\mathrm{FM}}$, 
solid circles). 
Negative values indicate a tendency to the antiparallel 
alignment while the positive values indicate a tendency to the parallel
alignment of local magnetic moments.
(b) Exchange parameter $J_{0}$ in the reference DLM configuration 
defined as the sum of corresponding exchange interactions $J_{0i}$ 
in the Fe-overlayer (crosses).
}
\end{figure}

In Fig.~\ref{j0} we compare the difference between total energies of the DLM and
FM phases with the calculated exchange parameter $J_{0}$,
Eq.~(\ref{J0-sum}), for the DLM reference state adopted in this paper.
Overall good agreement between both dependences over the whole
concentration range should be pointed out.
This confirms that the DLM state is a reasonable reference state
for the estimate of exchange integrals.
The small deviations are due to approximations inherent to the 
Heisenberg model (fixed size of local moments, the neglected 
contribution of induced moments in the substrate).

The observed concentration trend of the first NN interaction
and of the relative stability of the FM and DLM states can be
understood qualitatively on the basis of general properties of
a broad class of physical quantities in tight-binding models as
functions of band filling. \cite{r_1981_hsn, r_1983_hs}
These quantities include, e.g., total-energy differences,
local and non-local susceptibilities, magnetic and chemical
interactions, etc.; all of them change their sign due to
variation of the Fermi energy across the $d$-band.
In the present case, the position of the Fermi level with respect
to the Fe $d$-band is directly controlled by the concentration of
the Ta-W alloy substrate.

\begin{figure}
\begin{center}
\includegraphics[width=12cm]{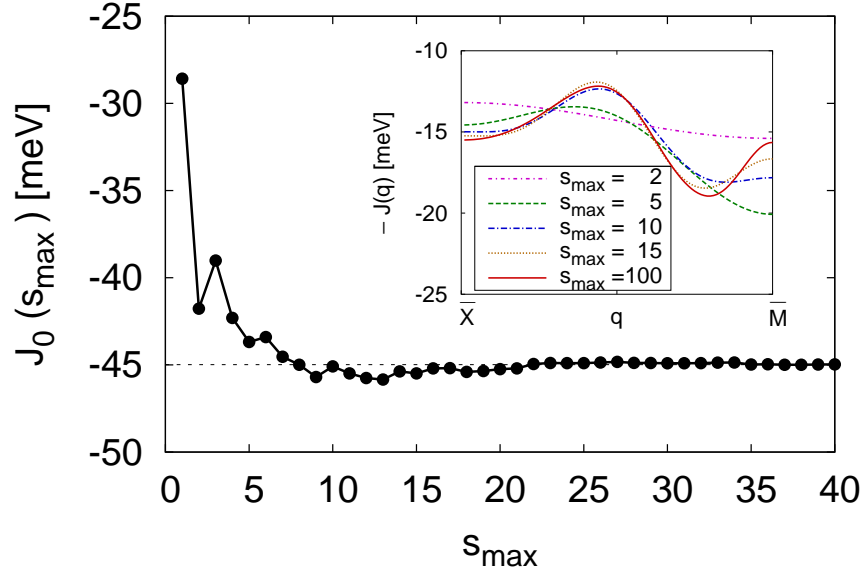}
\caption{\label{converg} (color online).
The partial sum of the exchange integrals, 
$J_{0}(s_{\rm max})=\sum_{s=1}^{s_{\rm max}} N_s J_s$ for the  
Fe$/$Ta$_{0.2}$W$_{0.8}$(001) system as a function of the shell 
number $s_{\rm max}$.  
$J_{s}$ is the exchange integral for the 
atomic shell $s$ and $N_{s}$ is the corresponding shell degeneracy, 
i.e., the number of equivalent atoms in a given shell.
The dependence of the lattice Fourier transform $-J({\bf q})$ 
(see Fig.~\ref{jq}(b)) on the number of shells $s_{\rm max}$ 
used in the Fourier transform is shown in the inset.
The last of the curves in the inset ($s_{\rm max}$=100) is essentially a converged result with respect to the shell number.
}
\end{center}
\end{figure}

Finally, in Fig.~\ref{converg} we illustrate the convergence of the 
exchange parameter $J_{0}$, Eq.~(\ref{J0-sum}), as a function 
of the number of shells included in the summation.
Note that now the shell degeneracies, $N_{s}$, are included.
Obviously, the inclusion of about 20 NN shells is sufficient for 
practical purposes (the deviation of the partial sum from the converged
value is less than 1~\%).
At least 10 shells have to be considered in order to predict the 
formation of an incommensurate magnetic ground state of 
Fe$/$Ta$_{0.2}$W$_{0.8}$(001) (see the next subsection for details). 
This is demonstrated in the inset of Fig.~\ref{converg}, where the 
lattice Fourier transform $J({\bf q})$ of the real-space 
exchange interactions $J_{0i}$ is magnified for the 
relevant wavevectors ${\bf q}$ in the surface Brillouin zone.
The position of the minimum of the spin-spiral energy, i.e., the wavevector  
${\bf q}$ for which $J({\bf q})$ has maximum, begins to approach the 
converged position for more than 10 shells while about 70 
shells are needed to reach the converged position with the precision of 
$0.1~\%$ of the wavevector ${\bf q}$.
The first two dominating NN interactions are obviously not enough 
to reproduce this local spin-spiral energy minimum.

\subsection{Magnetic stability}

Let us consider a class of possible arrangements (the so-called ${\bf q}-$waves)
of magnetic moments defined as
\begin{equation}\label{q-wave}
{\bf e}_i = (\sin \theta_i \cos \phi_i, \sin \theta_i \sin \phi_i, \cos \theta_i) \, ,
\end{equation}
where the polar angle $\theta_i = \theta_0$ is a constant and the azimuthal angle
$\phi_i={\bf q} \cdot {\bf R}_i$ is a function of the position vector ${\bf R}_i$
of site $i$.
The energy per lattice site corresponding to the HH (\ref{HH-def})
$E({\bf q},\theta_0)=-J({\bf q}) \sin^2 \theta_0 - J({\bf 0}) \cos^2 \theta_0$ is expressed in terms
of the lattice Fourier transform of the site-dependent exchange integrals
\begin{equation}\label{Jq-def}
J({\bf q}) = \sum_j \, J_{0j} \,
e^{i \, {\bf q} \cdot {\bf R}_{0j} } \, .
\end{equation}
It is easy to show that
\begin{equation}\label{Eqmin}
\min_{{\bf q},\theta_0} E({\bf q},\theta_0)= - \max_{\bf q} J({\bf q})
\end{equation}
and that $\theta_0 = \pi/2$ if the minimum of energy is achieved for ${\bf q} \neq {\bf 0}$.
Let us note that $J({\bf q})$ is closely connected to the energies of
magnetic excitations (magnons) in the system decribed by the HH, Eq.~(\ref{HH-def}).
The maximum of $J({\bf q})$, or, equivalently, the
minimum of $-J({\bf q})$, reached for a particular 
value of the vector ${\bf q}={\bf q}_0$ in the surface Brillouin 
zone indicates a tendency of the magnetic system to 
form a magnetic ground state characterized by 
that wave vector ${\bf q}_0$.
The wave vector ${\bf q}_0 = \bf 0$ (point $\bar{\Gamma}$ 
in the surface Brillouin zone) correponds to the FM ground state 
while a nonzero wave vector ${\bf q}_0$ corresponds to a more 
complex ground state including possible AFM or spin-spiral states.
For example, ${\bf q}_0$ at the point $\bar{M}$ of the surface 
Brillouin zone corresponds to the $c(2\times 2)$ (checkerboard-like) 
AFM ground state and ${\bf q}_0$ at the point $\bar{X}$ corresponds 
to the $p(2\times 1)$ (row-wise) AFM ground state.
A minimum of $-J({\bf q})$ outside any point of high-symmetry 
in the surface Brillouin zone suggests a tendency to form an incommensurate 
(spin-spiral) magnetic ground-state structure.

In this way, we are able to investigate the stability of a much broader 
class of magnetic systems as compared to the conventional total energy search.
It should be noted, however, that possible more general ground
states are not included in the present search and a more advanced
search based on the overlayer Heisenberg Hamiltonian,
Eq.~(\ref{HH-def}), would be needed.
Finally, we mention that in the present study we neglect the anisotropic part of the effective magnetic Hamiltonian,
i.e., the anisotropy that concerns orientation of the spins with respect to the underlying lattice.
Thus the relativistic effects (the magnetic anisotropy and the Dzyaloshinskii-Moriya interactions
(DMI)) as well as the magnetostatic (magnetic dipole-dipole) interactions are missing in the present analysis.
The dipole-dipole pair interactions between the first and second NN Fe-Fe pairs are at least two orders
of magnitude weaker (below 0.01 meV) than the typical exchange interactions in the present case.
However, their long range contributes significantly to formation of domain walls
and it might influence other details of the magnetic order too.
It is also known that the DMI can lead to long-range chiral structures 
due to weakening of FM-interactions, as it happens in the 
Mn$/$W(110) \cite{mnw110} and Mn$/$W(001) \cite{mnw001} systems.
We note that in a consistent theoretical treatment all possible anisotropic terms would have to be included
on the same footing.

\begin{figure}
\begin{center}
\includegraphics[width=12cm]{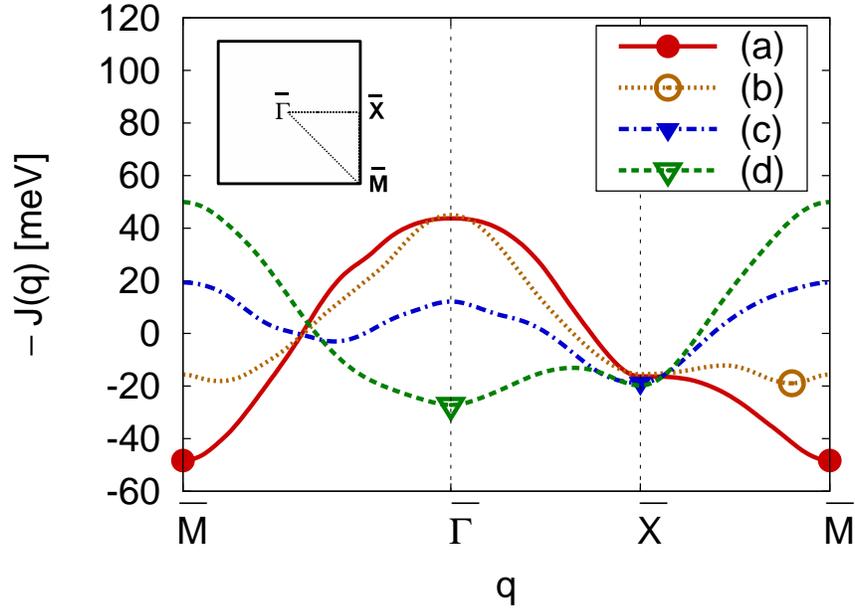}
\end{center}
\caption{\label{jq} (color online).
The lattice Fourier transform $J({\bf q})$ of exchange integrals 
in the Fe overlayer as defined by Eq.~(\ref{Jq-def}) along a chosen path
in the first surface Brillouin zone plotted for different substrate 
alloy compositions: (a) bcc Fe$/$W(001), (b) bcc Fe$/$Ta$_{0.2}$W$_{0.8}$(001),
(c) bcc Fe$/$Ta$_{0.5}$W$_{0.5}$(001), and (d) bcc Fe$/$Ta(001).
The minimum of $-J({\bf q})$ for each substrate composition is indicated
by a symbol on the corresponding curve.
The inset shows the first surface Brillouin zone and the corresponding
high-symmetry points.
}
\end{figure}

The curves of $-J({\bf q})$ plotted in Fig.~\ref{jq} and 
especially the positions of their minima reflect the dependence of 
the magnetic ground state on the composition of the substrate.
A clearly pronounced minimum for Fe$/$W(001) at the $\bar{M}$ point
in the surface Brillouin zone confirms the total energy search,
namely that the checkerboard $c(2\times 2)$-AFM is the ground state 
of the system.
With increasing Ta content, the stability of the $c(2\times 2)$-AFM state
weakens and for about 20~\% of Ta atoms a weak minimum corresponding 
to an incommensurate state (with the wavevector on the line 
$\bar{X}-\bar{M}$ close to  $\bar{M}$) develops.
It should be noted that this minimum is sensitive to the number of
shells included in the evaluation of $-J({\bf q})$ 
and is missing if we consider only few leading exchange interactions 
(see previous subsection).
We cannot exclude the possibility, however, that this shallow local minimum will 
disappear when the relativistic effects are included.
The magnetocrystalline anisotropy, for example, strengthens the tendency
towards collinear AFM alignment and competes with the DMI interaction 
which promotes the rotating, non-collinear magnetism as it was 
demonstrated recently for the Fe-double chains on the fcc-Ir(001) surface
\cite{relat}.
If the Ta content further increases, the $p(2\times 1)$-AFM configuration
becomes the ground state as it is illustrated for the case with
50~\% of Ta atoms 
where the minimum of $-J({\bf q})$ occurs at 
$\bar{X}$, see Fig.~\ref{jq}.
In the Ta-rich alloy the stability of the FM phase increases as
compared to the $c(2\times 2)$ one until for Fe$/$Ta(001) the wave 
vector at $\bar{\Gamma}$ (or the FM state) becomes the magnetic
ground state.

The observed tendency to a non-collinear ground state for $x \approx 0.2$ 
can be explained by frustration effects on the square Fe lattice accompanying
the dominating first and second nearest-neighbor interactions that are both 
negative for concentrations $x \leq 0.45$, see Fig.~\ref{j5}.
For W-rich substrates ($x \to 0$), the first nearest-neighbor interaction 
dominates ($|J_1| \gg |J_2|$) and the $c(2\times 2)$ AFM ground state is not 
frustrated. 
For equiatomic concentrations ($x \to 0.5$), the second nearest-neighbor 
interaction is the strongest one ($|J_2| \gg |J_1|$) and the $p(2\times 1)$ AFM 
ground state is not frustrated either.
However, for compositions around $x \approx 0.25$, both interactions are of 
comparable magnitudes ($J_1 \approx J_2$) which leads to pronounced frustration 
of these simple AFM states. 
Moreover, these AFM states become nearly degenerate at
$x=0.2$, where the two leading interactions satisfy roughly a relation
$J_1 = 2 J_2$. \cite{prl-stefan}
The formation of a single-${\bf q}$ spin spiral for $x \approx 0.2$
(or of a more complex spin
arrangement not considered here) represents thus a natural consequence of
existing interactions in the Fe-monolayer.

Finally, the present results (e.g., the critical concentration
for the AFM to FM crossover) can be influenced quantitatively
by possible structural changes in real system.
A recent study \cite{taw_bulk} indicates a possibility for
the B2-ordering in a bulk bcc-TaW alloy over a broad concentration
range.
Also, a smaller surface energy of bcc-Ta as compared to the bcc-W
\cite{surfen} together with the larger size of Ta atoms seems to
indicate a possibility for the segregation of Ta-atoms at the
bcc-(Ta,W)(001) alloy surface.
On the other hand, the qualitative conclusions of the present
paper remain unchanged.

\section{Conclusions}

We have developed a new approach to study the magnetic phase
stability of magnetic overlayers on non-magnetic substrates.
The approach consists in the evaluation of exchange integrals
between local magnetic moments
in the magnetic overlayer using the adiabatic
approximation and the real-space Green function approach.
This allows us to use this approach also for magnetic overlayers
on a random substrate and/or for random alloy magnetic overlayers.
Estimated exchange interactions between pairs of 
local moments in the overlayer are used to construct the effective two-dimensional
Heisenberg Hamiltonian, whose stability with respect to periodic
spin excitations is investigated.
The maxima of the lattice Fourier transform of the site-dependent exchange
integrals are searched for as indications of stable periodic spin structures.
Such a search allows one to investigate the stability of a much
broader class of possible magnetic configurations as compared
to the conventional total energy search which is limited to a few
empirically chosen configurations.
The present approach can be extended in future to include more general magnetic configurations
and/or anisotropic effects (spin-orbit and dipole-dipole interactions);
the latter are indispensable for estimation of the corresponding Curie/Neel temperatures.
As a case study we have investigated in detail the magnetic phase
stability of the Fe overlayer on the bcc-Ta$_{x}$W$_{1-x}$(001) 
random substrate. 
The exchange interactions for this system were extracted from 
the DLM state, which involves no correlation among spins of overlayer atoms.
Our results are in good agreement with the results of a recent study 
based on the total energy search and confirm basic approximations 
adopted in this study.
In addition, we have predicted a possible incommensurate magnetic 
configuration for the W-rich substrate alloy (at about 20~\% of Ta).\\

\begin{acknowledgments}
The research was carried out within the projects AV0Z10100520,
AV0Z10100521, and AV0Z20410507 of the Academy of Sciences of 
the Czech Republic.
Financial support was provided by the Grant Agency of the Academy 
of Sciences of the Czech Republic (Project A100100616) and 
Czech Science Foundation (Projects 202/07/0456 and 202/09/0775).
O.~B.\ would like to acknowledge funding support from the ANR of France, Grant No.
ANR-06-NANO-053-01.
\end{acknowledgments}

\appendix*

\section{Method of infinitesimal rotations in the DLM state}

The purpose of this Appendix is to show a close relation of
the formula (\ref{Jij-dlm}) for exchange interactions in the DLM-reference
state to the well-known formula (\ref{Jij-fm}) valid for any collinear
reference state, including the FM state.
The latter formula for the pair interaction $J_{ij}$ between local
moments on lattice sites $i$ and $j$ can be written as
\begin{equation}
J_{ij} = \frac{1}{4\pi} \,
\mathrm{Im} \int_{C} w_{ij}(z) \, \mathrm{d}z ,
\label{eqa_jij}
\end{equation}
where the integrated function in the TB-LMTO method has a form
\begin{equation}
w_{ij} = \mathrm{tr}_{L} \left\{
\left( P^{\uparrow}_{i} - P^{\downarrow}_{i} \right)
\bar{g}^{\uparrow}_{ij}
\left( P^{\uparrow}_{j} - P^{\downarrow}_{j} \right)
\bar{g}^{\downarrow}_{ji}
\right\} ,
\label{eqa_wij_fm}
\end{equation}
where the energy argument $z$ of each quantity has been omitted.
Here we assume for simplicity that possible chemical disorder (random
occupation by different species) is confined to non-magnetic sites
whereas the sites $i$ and $j$ in (\ref{eqa_jij}, \ref{eqa_wij_fm})
are occupied by a single species with non-zero local moment.

The standard treatment of the DLM state employs the CPA \cite{r_1985_gps}
which leads to the alloy analogy, i.e., the electronic structure can be
obtained from a binary equiconcentration random alloy A$_{0.5}$B$_{0.5}$
with collinear spin structure,
where the spin-dependent potential functions of both components are
given by $P^{A\sigma}_{i} = P^{B-\sigma}_{i} \equiv P^{\sigma}_{i}$
(where $\sigma = \uparrow, \downarrow$, and $-\!\uparrow = \downarrow$,
$-\!\downarrow = \uparrow$).
The symbol A (B) represents atoms with local moments pointing upwards
(downwards); for symmetry reasons, the configurationally averaged
Green function as well as the coherent potential functions are
spin-independent, i.e., $\bar{g}^{\sigma}_{ij} \equiv \bar{g}_{ij}$
and $\mathcal{P}^{\sigma}_{i} \equiv \mathcal{P}_{i}$.

Exchange interactions in the DLM-reference state can naturally be
obtained from energy changes accompanying infinitesimal rotations of
local moments that were originally pointing in the same direction.
This corresponds to occupation of both sites $i$ and $j$ by
the same atomic species, e.g., by A atoms.
The configuration average of the exchange interaction $J_{ij}$
leads to its original form (\ref{eqa_jij}) with the function
$w_{ij}(z)$ replaced by
\begin{equation}
w_{ij} = \mathrm{tr}_{L} \left\{
\left( P^{{\rm A}\uparrow}_{i} - P^{{\rm A}\downarrow}_{i} \right)
\bar{g}^{{\rm A,A}\uparrow}_{ij}
\left( P^{{\rm A}\uparrow}_{j} - P^{{\rm A}\downarrow}_{j} \right)
\bar{g}^{{\rm A,A}\downarrow}_{ji}
\right\} ,
\label{eqa_wij_dlm_s}
\end{equation}
where $\bar{g}^{{\rm A,A}\sigma}_{ij}$ denotes the conditionally
averaged Green function in spin channel $\sigma$ between sites
$i$ and $j$ occupied by atoms A.
Note that the conditionally averaged Green functions depend on the
spin index.
Their values within the CPA and for $i \neq j$ are given by
\begin{equation}
\bar{g}^{{\rm A,A}\sigma}_{ij} =
\tilde{f}^{{\rm A}\sigma}_{i}
\, \bar{g}_{ij} \,
f^{{\rm A}\sigma}_{j} ,
\label{eqa_gij}
\end{equation}
where the spin-dependent factors on the r.h.s.\ can be expressed in
terms of the spin-independent on-site Green functions $\bar{g}_{ii}$
and coherent potential functions $\mathcal{P}_{i}$ as
\begin{eqnarray}
f^{{\rm A}\sigma}_{i} & = &
\left[ 1 + \left( P^{\sigma}_{i} - \mathcal{P}_{i} \right)
\bar{g}_{ii} \right]^{-1} \equiv f^{\sigma}_{i} ,
\nonumber\\
\tilde{f}^{{\rm A}\sigma}_{i} & = &
\left[ 1 + \bar{g}_{ii} \left( P^{\sigma}_{i} - \mathcal{P}_{i}
\right) \right]^{-1} \equiv \tilde{f}^{\sigma}_{i} .
\label{eqa_ftf}
\end{eqnarray}

The scattering due to the atomic species A and B with
respect to the effective DLM-medium is described by spin-dependent
single-site t-matrices
$t^{A\sigma}_{i} = t^{B-\sigma}_{i} \equiv t^{\sigma}_{i}$, where
\begin{equation}
t^{\sigma}_{i} = f^{\sigma}_{i}
\left( P^{\sigma}_{i} - \mathcal{P}_{i} \right)
= \left( P^{\sigma}_{i} - \mathcal{P}_{i} \right)
\tilde{f}^{\sigma}_{i} .
\label{eqa_tmat}
\end{equation}
Their difference can be related to the difference of the potential
functions $P^{\sigma}_{i}$, namely,
\begin{eqnarray}
t^{\uparrow}_{i} - t^{\downarrow}_{i} & = &
\left( P^{\uparrow}_{i} - \mathcal{P}_{i} \right)
\tilde{f}^{\uparrow}_{i}
- f^{\downarrow}_{i}
\left( P^{\downarrow}_{i} - \mathcal{P}_{i} \right)
\nonumber\\
 & = &
 f^{\downarrow}_{i} \left\{ \left[
 1 + \left( P^{\downarrow}_{i} - \mathcal{P}_{i} \right)
 \bar{g}_{ii} \right]
 \left( P^{\uparrow}_{i} - \mathcal{P}_{i} \right)
 \right.
\nonumber\\
 & & \left. { }
 - \left( P^{\downarrow}_{i} - \mathcal{P}_{i} \right)
\left[ 1 + \bar{g}_{ii} \left( P^{\uparrow}_{i} -
\mathcal{P}_{i} \right) \right]
\right\} \tilde{f}^{\uparrow}_{i}
\nonumber\\
 & = & f^{\downarrow}_{i}
 \left( P^{\uparrow}_{i} - P^{\downarrow}_{i} \right)
 \tilde{f}^{\uparrow}_{i} ,
\label{eqa_tdif1}
\end{eqnarray}
and, similarly,
\begin{equation}
t^{\uparrow}_{i} - t^{\downarrow}_{i} =
  f^{\uparrow}_{i}
 \left( P^{\uparrow}_{i} - P^{\downarrow}_{i} \right)
 \tilde{f}^{\downarrow}_{i} .
\label{eqa_tdif2}
\end{equation}
Substitution of Eq.~(\ref{eqa_gij}) into Eq.~(\ref{eqa_wij_dlm_s}),
cyclic invariance of the trace and identities
(\ref{eqa_tdif1}, \ref{eqa_tdif2}) lead to the final expression
of the function $w_{ij}(z)$ in the pair interaction $J_{ij}$
(\ref{eqa_jij}):
\begin{equation}
w_{ij} = \mathrm{tr}_{L} \left\{
\left( t^{\uparrow}_{i} - t^{\downarrow}_{i} \right)
\bar{g}_{ij}
\left( t^{\uparrow}_{j} - t^{\downarrow}_{j} \right)
\bar{g}_{ji}
\right\} .
\label{eqa_wij_dlm_f}
\end{equation}
This result is equivalent to the formula (\ref{Jij-dlm}) in the
main text.

%%\bibliography{ftw}
%%%%%%%%%%%

%%%%%%%%%%%

\newpage

\newpage

\newpage

\newpage

\newpage

\end{document}